\documentclass[11pt,dvips]{article}
\usepackage{amscd,verbatim}
\usepackage[all]{xy}
\input xypic
\usepackage{graphicx}
\usepackage{amsmath}
\usepackage[psamsfonts]{amssymb}
\usepackage{amsthm}
\usepackage{euscript}

\usepackage{latexsym}

\renewcommand{\(}{\begin{equation}}
\renewcommand{\)}{end{equation} \vspace{-.05in}\linebreak}

\newcounter{saveeqn}
\newcounter{savealpheqn}

\newcommand{\alpheqn}{\setcounter{saveeqn}{\value{equation}}%
  \stepcounter{saveeqn}\setcounter{equation}{0}%
  \renewcommand{\theequation}{\mbox{\arabic{section}.\arabic{saveeqn}
\alph{equation}}}
  \renewcommand{\)}{\end{equation}}}
\def\part#1{\frac{\partial}{\partial{#1}}}%
\def\group#1{\refstepcounter{equation}\setcounter{saveeqn}{\value{equati
on}}%
  \label{#1}\setcounter{equation}{0}%
\renewcommand{\theequation}{\mbox{\arabic{section}.\arabic{saveeqn}
\alph{equation}}}
  \renewcommand{\)}{\end{equation}}}
\newcommand{\reseteqn}{\setcounter{equation}{\value{saveeqn}}%
  \renewcommand{\theequation}{\arabic{section}.\arabic{equation}}%
  \renewcommand{\)}{\end{equation}}}

\newcommand{\aalpheqn}{\setcounter{saveeqn}{\value{equation}}%
  \stepcounter{saveeqn}\setcounter{equation}{0}%
  \renewcommand{\theequation}{\mbox{
        \Alph{subsection}.\arabic{saveeqn}\alph{equation}}}
   \renewcommand{\)}{\end{equation}}}
\newcommand{\areseteqn}{\setcounter{equation}{\value{saveeqn}}%
  \renewcommand{\theequation}{\Alph{subsection}.\arabic{equation}}%
  \renewcommand{\)}{\end{equation}}}

\renewcommand{\thefootnote}{\alph{footnote}}
\renewcommand{\(}{\begin{equation}}
\renewcommand{\)}{\end{equation}}
\newcommand{\ba}{\begin{eqnarray}}
\newcommand{\ea}{\end{eqnarray}}

\renewcommand{\r}{\rho}
\newcommand{\bp}{\mathop{\vtop{\ialign{##\crcr
   $\hfil\displaystyle{}\hfil$\crcr\noalign{\kern-13pt\nointerlineskip}
   \BIG{(}\hskip0pt\crcr\noalign{\kern3pt}}}}}
\newcommand{\cbp}{\mathop{\vtop{\ialign{##\crcr
   $\hfil\displaystyle{}\hfil$\crcr\noalign{\kern-13pt\nointerlineskip}
   \BIG{)}\hskip0pt\crcr\noalign{\kern3pt}}}}}
\newcommand{\pa}{\mathop{\vtop{\ialign{##\crcr

$\hfil\displaystyle{\oplus}\hfil$\crcr\noalign{\kern+1pt\nointerlineskip
}
   \hspace{.08in}$^{\alpha=0}$\hskip6pt\crcr\noalign{\kern3pt}}}}}

  {\renewcommand{\theequation}{\Alph{subsection}.\arabic{equation}}%
   \renewcommand{\thesubsection}%
                {Appendix \Alph{subsection}.\setcounter{equation}{0}}%
   \renewcommand{\alpheqn}{\aalpheqn}%
   \renewcommand{\reseteqn}{\areseteqn}
   }

\newcommand{\R}{\ensuremath{\mathbb R}}

\newcommand{\C}{\ensuremath{\mathbb C}}

\newcommand{\Q}{\ensuremath{\mathbb Q}}

\newcommand{\Z}{\ensuremath{\mathbb Z}}

\def\r{\rightarrow}

\newcommand{\beq}{\begin{equation}}
\newcommand{\beg}[2]{\begin{equation}\label{#1}#2\end{equation}}
\newcommand{\eeq}{\end{equation}}

\numberwithin{equation}{section}

\def\hsp#1{\hspace{#1in}}

\newcommand{\rref}[1]{(\ref{#1})}

\catcode`\@=11
\def\vereq#1#2{\lower3pt\vbox{\baselineskip1.5pt \lineskip1.5pt
\ialign{$\m@th#1\hfill##\hfil$\crcr#2\crcr\sim\crcr}}}
\catcode`\@=12

\makeatletter
\newcommand\figcaption{\def\@captype{figure}\caption}
\newcommand\tabcaption{\def\@captype{table}\caption}
\makeatother
\renewcommand{\(}{\begin{equation}}
\renewcommand{\)}{\end{equation}}

\newcommand{\CC}{{\mathbb C}}

\theoremstyle{plain}

\theoremstyle{definition}

\newcommand{\CP}{\CC \text{P}}

\begin{document}

\newcommand{\bea}{\begin{eqnarray}}
\newcommand{\eea}{\end{eqnarray}}

\begin{titlepage}
\begin{flushright}


hep-th/0501060
\end{flushright}

\vspace{2em}
\def\thefootnote{\fnsymbol{footnote}}

\begin{center}
{\Large\bf Type II string theory and modularity} \footnote{I. K.
is supported by NSF grant DMS 0305853, and H. S. is supported by
the Australian Research Council.}
\end{center}
\vspace{1em}
\begin{center}
Igor Kriz\footnote{E-mail: \tt ikriz@umich.edu}$^1$ and Hisham
Sati \footnote{E-mail: \tt hsati@maths.adelaide.edu.au}$^{2,3}$

\end{center}

\begin{center}
\vspace{1em}
{\em  { $^1$Department of Mathematics\\
            University of Michigan\\
            Ann Arbor, MI 48109,\\
            USA\\
\hsp{.3}\\
$^2$Department of Physics\\
       University of Adelaide\\
       Adelaide, SA 5005,\\
       Australia}\\
\hsp{.3}\\
$^3$Department of Pure Mathematics\\
       University of Adelaide\\
       Adelaide, SA 5005,\\
       Australia}\\

\end{center}

\vspace{0em}

\begin{abstract}
This paper, in a sense, completes a series of three papers. In the
previous two \cite{KS} \cite{KSB}, we have explored the
possibility of refining the $K$-theory partition function in type
II string theories using elliptic cohomology. In the present
paper, we make that more concrete by defining a fully quantized
free field theory based on elliptic cohomology of $10$-dimensional
spacetime. Moreover, we describe a concrete scenario how this is
related to compactification of F-theory on an elliptic curve
leading to IIA and IIB theories. We propose an interpretation of
the elliptic curve in the context of elliptic cohomology. We
discuss the possibility of orbifolding of the elliptic curves and
derive certain properties of F-theory. We propose a link of this
to type IIB modularity, the structure of the topological
Lagrangian of M-theory, and Witten's index of loop space Dirac
operators. The discussion suggests a $S^1$-lift of type IIB and an
F-theoretic model for type I obtained by orbifolding that for type
IIB.


\end{abstract}

\vfill

\end{titlepage}
\setcounter{footnote}{0}
\renewcommand{\thefootnote}{\arabic{footnote}}
\newcommand{\cform}[3]{\begin{array}{c}
{\scriptstyle #3}\\
#1\\
{\scriptstyle #2}\end{array}}

\pagebreak

\renewcommand{\thepage}{\arabic{page}}

\section{Introduction}

The main purpose of this paper is to complete, at least in some
sense, our investigation \cite{KS} \cite{KSB} of a refinement of
the $K$-theory RR partition function of type II string theory
\cite{MW} \cite{DMW}. The RR $K$-theory partition function is
obtained from the observation that $K^0(X)$ and $K^1(X)$
classifies the RR sources in type IIA and IIB string theory on a
$10$-manifold $X$ \cite{Wi1}\cite{Petr}. The $K$-theory partition
function is a theta function obtained from quantizing essentially
the free field theory on those sources, i.e. where the Lagrangian
is essentially the Hermitian metric on a field strength which is
set up in such a way that the phase term of the Lagrangian can be
thought of as an index (see \cite{DMW}). The main result of
\cite{DMW} is that in type IIA string theory, this partition
function agrees with the partition function of M-theory
compactified on $S^1$ where the effective action is taken to be
the Chern-Simons term together with correct normalization and a
$1$-loop correction term, which makes the phase again an index.
This time, however, the index is a combination of an $E_8$-index
and a Rarita-Schwinger index on a $12$-manifold $Z^{12}$ which
cobords $X\times S^1$. An extension of part of the construction to
twisted K-theory, i.e. to include NSNS background, was studied in
\cite{HM}.

\vspace{3mm} In \cite{KS}, we observed that an anomaly $W_7(X)$
detected in \cite{DMW} on both the IIA and M-theory sides
coincides with the anomaly of orientation with respect to elliptic
cohomology. This led us to propose in \cite{KS} a refinement of
the $K$-theory partition function, which would be based on
elliptic cohomology. When fully quantizing the theory associated
with that partition function, we encountered a refinement of the
obstruction $W_7(X)$ to $w_4(X)$ (one has $W_7(X)=\beta
Sq^2(w_4(X))$. This suggests that the elliptic partition function
must be related to a scenario where type II string theory is
unified with type I and heterotic, where a $4$-dimensional
obstruction is detected -- such obstruction is not known in type
II theory per se.

\vspace{3mm} Elliptic cohomology is a certain refinement of
$K$-theory which is introduced in topology, and which has the
striking property that its coefficients (homotopy groups, or
cohomology groups of a point) consist of modular forms, of weight
$k/2$ where $k$ is dimension. This led us to propose in
\cite{KSB}, after eliminating some simpler scenarios, that the
elliptic cohomology partition function may be a step toward
solving the puzzle of \cite{DMW} related to IIB modularity: when
one writes the $K$-theory partition function for type IIB, it does
not seem to accomodate a modularity in the presence of an $H_3$
source. It is remarked in \cite{DMW} and further investigated in
\cite{KSB} that twisted $K$-theory, which is the first approach
which comes to mind, does not solve the problem.

\vspace{3mm} The papers \cite{KS} \cite{KSB} left unanswered the
question where the elliptic cohomology source partition function
of type II string theory really comes from. In \cite{DMW}, the IIA
partition function is linked to M-theory compactified on a circle.
What, if any, is the analogous link for the elliptic cohomology
partition function? In this paper, we attempt to answer that
question, and derive some implications from the answer.

\vspace{3mm} The scenario we propose is that the elliptic
cohomology partition function is related to compactification of
F-theory on an elliptic curve $E$, which is a theory first
suggested by Vafa \cite{Vafa}. We propose that the modularity of
elliptic cohomology, which makes the partition function itself
modular, comes from modularity in $H^1(E)$, i.e. from the moduli
parameter of the elliptic curve $E$. This is in fact somewhat
linked to the purely mathematical paper \cite{ell}, in which it
was derived that in developing a purely mathematical link between
elliptic cohomology and conformal field theory, modularity of
elliptic cohomology is related to an elliptic curve in spacetime,
not merely to a genus one worldsheet.

\vspace{3mm} In the present paper, the link in fact goes one step
further: in the same way in which the action of the free
approximation to the RR sector of type II theory is related to the
$\hat{A}$-genus, the elliptic cohomology form of the RR sector of
the theory is related to the Witten genus. Why is that? We propose
an explanation. In \cite{W-loop}, Witten shows that the Witten
genus is related to index of elliptic operators on loop space.
Using a standard lifting \cite{Sag} of the M-theory action to
F-theory \cite{Vafa} (cf. \cite{Pope} for the reduction), we
propose that in F-theory compactified on a circle, the phase
factor analogous to that analyzed in \cite{DMW} may be obtained as
loop versions of the $E_8$ and Rarita-Schwinger indices. We view
the loop group bundles \cite{Allan} \cite{J1} \cite{HM} as coming
from bundles over the loop space of spacetime. Thus, we possibly
link modularity in four different places: S-duality in type IIB
string theory, elliptic cohomology (spacetime aspect), the Witten
genus and F-theory fiber.

\vspace{3mm} There is one caveat to our story: The modularity of
elliptic cohomology is not entirely anomaly-free, and accordingly
the modular forms are really just automorphic forms of a certain
level ($=3$ if we focus on integrality at the prime $2$). There is
a theory constructed by Mike Hopkins and known as $TMF$
(topological modular forms, the connective form of the theory is
known as $tmf$) which remedies the difficulty: this theory has a
complete anomaly-free modularity. However, the price for that is
that it is again a much more complicated generalized cohomology
theory, which can no longer be called an elliptic cohomology
theory -- it is obtain from elliptic cohomology theory by a
procedure which we could compare to the physical procedure of
orbifolding.

\vspace{3mm} Indeed, this orbifolding seems to correspond to
orbifolding in physics literally. Orbifolding type IIB in the
worldsheet sense (reversing chiralities) leads to type I string
theory. Orbifolding $S^1$-compactified M-theory with respect to
the $11$-th dimension in spacetime leads to Ho\v{r}ava-Witten
M-theory \cite{Hora}. We deduce from this a relation between this
type of worldsheet and spacetime orbifolding. In F-theory, we
further predict a more complicated orbifolding with respect to
(roughly) the group $SL(2,\Z/2)$, which should produce an ``ideal
F-theory'' governed by $TMF$, which would have a complete, anomaly
free modularity. We do not however work out this optimal scenario
in detail, and in most of this paper still use just ordinary
elliptic cohomology instead.

\vspace{3mm} The Chern-Simons part of the action of M-theory was
used by Witten and rewritten as in a symmetric way as a cubic
expression in the four-form \cite{flux}. One wonders whether there
is a general mathematical reason for such a structure, beyond just
being able to use some form of Stokes' theorem. We show that the
lifted Chern-Simons term can be written as a Massey triple product
and the one-loop term can be explained as being a part of the
Massey product indeterminacy.

\vspace{3mm} The Lagrangian of the ultimate twelve dimensional
theory is not completely worked out in the present paper. There
are at least two sources of topological terms in such Lagrangian,
one of them which should be related to M-theory upon
compactification on $S^1$, another which should be related to
M-theory by cobordism. However, it is possible that a Massey
product device similar to the one mentioned above can also be used
to unify these situations -- we make a comment to that effect.

\vspace{3mm} The present paper is organized as follows: In Section
\ref{s2}, we revie basic features of IIB modularity from a
classical and quantum-mechanical point of view. In Section
\ref{stheor}, we review the Lagrangians of known theories in $12$
dimensions. In Section \ref{s4}, we present our evidence for
topological modular forms from the IIB modularity question and
also from the cobordism approach to M-theory \cite{flux}. In
Section \ref{s5}, we give our main explanations about the elliptic
cohomology partition function and its relation to F-theory and the
Witten genus. Finally, in Section \ref{s6}, we give some general
comments on what would be needed to discuss F-theory at physical
signatures. In order to make the paper more self-contained and
more accessible to physicists, we included a brief appendix on
topological modular forms.

\section{Type IIB and modularity}
\label{s2} In this section we review the basic features of type
IIB supergravity and string theory that will be relevant for our
later discussions.

\vspace{3mm} The bosonic field content of type IIB supergravity
is: metric $g$, two scalars $\phi$ and $\chi$, a complex 3-form
field strength $G_3$ and a real self-dual five-form field strength
$F_5$. The fermionic content is: two gravitini $\psi^i$ ($i=1,2$)
of the same chirality, i.e. sections of
$S(X)^{\pm}\otimes(TX-2{\mathcal O})$ (with the same choice of
sign), and two dilatini of the opposite chirality to the
gravitini, i.e. $\lambda^i \in \Gamma[S(X)^{\mp}]$. The two
scalars parametrize an upper half plane ${\mathcal
H}=SL(2,\Bbb{R})/U(1)$. In a fixed $U(1)$ gauge, the global
$SL(2,\Bbb{R})$ induces on the fields, collectively $\Phi$, a
$U(1)$ transformation that depends on their $U(1)$ charge
$q_{\Phi}$, as \cite{Seth} \( \Phi \longrightarrow \Phi
\left(\frac{c{\overline \tau}+d}{c\tau +d}
  \right)^{\frac{q_{\Phi}}{2}}.
\)

\vspace{3mm} The $SL(2,\Bbb{R})$ symmetry is broken down to the
local discrete subgroup $SL(2,\Bbb{Z})$ by nonperturbative quantum
effects. The arithmetic subgroup is conjectured to be an exact
symmetry of type IIB string theory. Its action factorizes into a
projective action on the complex scalar $\tau$ and a
charge-conjugation that reverses the signs of the two 2-forms and
leaves $\tau$ invariant.

\vspace{3mm} Strings with fractional charge do not exist and so
the type IIB string must be $SL(2,\Bbb{Z})$ invariant \cite{HT}
\cite{Var}. The action can be written in a manifestly
$SL(2)$-invariant way as
\begin{eqnarray}
S_{IIB}&=&\frac{1}{2\kappa_{10}^2} \int_{X^{10}} d^{10}x \sqrt{-g}
\left[ {\mathcal R} -\frac{1}{4} Tr(\partial {\mathcal M}
{\mathcal M}^{-1})^2 -\frac{1}{12}{\bar H}_{\mu \nu \rho}^T
{\mathcal M} {\bar H}^{\mu \nu \rho} - \frac{1}{4}{\tilde F}_{5}^2
\right]
\nonumber\\
&-&\frac{1}{8\kappa_{10}^2} \int_{X^{10}} C_4 \wedge H^i\wedge H^j
\epsilon_{ij},
\end{eqnarray}

where ${\mathcal M}$ is the metric on the coset $SL(2)/U(1)$ (i.e.
the upper half plane) given by \( {\mathcal M}=\frac{1}{Im(\tau)}
\left(\begin{array}{cc} |\tau|^2 & Re(\tau) \\
                Re(\tau) & 1 \end{array} \right),
\) ${\bar H}=\left( \begin{array}{c} H^1 \\ H^2 \end{array}
\right)$ is the doublet of three-forms with $H^i=(H_3,F_3)$. The
five-form ${\tilde F_5}$ is an RR field strength modified by the
RR and NS three-forms, i.e. \( {\tilde F_5}=F_5
-\frac{1}{2}C_2\wedge H_3 + \frac{1}{2}B_2\wedge F_3. \label{F5}
\) with ${F}_5=dC_4$, which can be written as \( {\tilde F_5}=F_5
+ \frac{1}{2}\epsilon_{ij} B^i\wedge H^j , ~~ i,j=\{1,2\}, \) and
$C_4$ is the $SL(2)$-invariant RR 4-form potential. This way we
see that the two scalars, namely the dilaton from the NS sector
and the axion from the RR sector, can be viewed as the coordinates
on the upper half-plane. So the modular parameter is built out of
the dilaton $\phi$ and the axion (= $0$-form RR potential),
$\tau=C_0+ie^{-\phi}$.

 \vspace{3mm}
The above action is invariant under the $SL(2,\Bbb{Z})$
transformations
\begin{eqnarray}
{\mathcal M}'&=&\Lambda {\mathcal M} \Lambda^T
\\
{\overline H}'&=&(\Lambda^T)^{-1} {\overline H}
\end{eqnarray}
with the metric in the Einstein frame being invariant, $g'_{\mu
\nu}=g_{\mu \nu}$, where the group element is $\Lambda=\left(
\begin{array}{cc} a & b \\ c&d \end{array} \right)$, with
$ad-bc=1$.

Alternatively, one can choose to use complex differential forms
and write the effective action of type IIB string theory in the
$SL(2,\Bbb{Z})$-invariant form (in the Einstein frame, see e.g.
\cite{Gid})

\begin{eqnarray}
S_{IIB}&=&\frac{1}{2\kappa_{10}^2}\int_{X^{10}} d^{10}x \sqrt{-g}
\left[{\mathcal R} - \frac{1}{2} \frac{\partial_M \tau \partial^M
{\bar \tau}} {(Im \tau)^2} -\frac{1}{12} {G_3\wedge *{\overline
G_3}} -\frac{1}{4} {\tilde F_5}^2  \right]
\nonumber\\
&+&\frac{1}{8i \kappa_{10}^2} \int_{X^{10}} {C_4\wedge G_3 \wedge
{\overline G_3}} + S_{(2p+1)-branes}.
\end{eqnarray}
Here the RR field strength $F_3$ and the NS field $H_3$ are
grouped as $SL(2,\Bbb{Z})$ doublet into the invariant complex
field \( G_3=\frac{1}{\sqrt{Im \tau}}(F_3 -\tau H_3) \label{eg3}
\) and similarly for the complex conjugate field, \( {\overline
G}_3=\frac{1}{\sqrt{Im \tau}}(F_3 -{\overline \tau} H_3). \)

The self-duality for ${\tilde F_5}$ cannot be seen at the level of
the above action but has to be imposed as an extra condition on
the equations of motion. The action is obviously invariant under
$SL(2,\Bbb{Z})$ transformations.

\vspace{3mm}
 The S-duality transformation is the subset of the above
$SL(2,\Bbb{Z})$ transformations given by $a=d=0$ and $b=-c=1$, so
that the fields transform as
\begin{eqnarray}
\tau &\rightarrow& -1/{\tau}
\nonumber\\
B_2 &\rightarrow& C_2
\nonumber\\
C_2 &\rightarrow& -B_2,
\end{eqnarray}
and again the metric and the five form are left invariant.

The moduli space of scalar fields is then $SL(2,\Bbb{Z})\backslash
{\mathcal H}$. The supersymmetry algebra has an automorphism
group, a continuous $U(1)$ R-symmetry that rotates the
supercharges, and this is broken down to a discrete subgroup
\cite{GG} $\Bbb{Z}_4=SL(2,\Bbb{Z})\cap U(1)$ that interchanges the
two supercharges and reverses the spatial worldsheet direction.

\vspace{3mm} The ${\Bbb{Z}}_4$ symmetry (see e.g. \cite{Berg})
generated by the elements $a=0$, $b=1$, $c=-1$, $d=0$, inverts the
modular parameter $\tau$ as $\tau \rightarrow \frac{-1}{\tau}$, so
that for vanishing axion $C_0=0$, this inverts the coupling
constant $e^{-\phi}\rightarrow e^{\phi}$, which can be interpreted
as the weak/strong coupling duality (S-duality). This
${\Bbb{Z}}_4$ symmetry also acts on the NS and RR 2-forms as
$B_2^{(1)} \rightarrow -B_2^{(2)}$ and $B_2^{(2)} \rightarrow
B_2^{(1)}$, so that $G_3$ and ${\bar G_3}$ are interchanged and
${\tilde F}_5$ is of course still invariant. This duality also
acts on the metric in the string frame, and that is why one has to
use the Einstein frame to get a duality-invariant action. Applying
${\Bbb{Z}}_4$ twice gives a ${\Bbb{Z}}_2$ with almost trivial
effect, in the sense that it leaves $\tau$ invariant but changes
the sign of the two 2-forms $B_2^{(i)}$.

\vspace{3mm} There is no one-loop correction in type IIB in ten
dimensions analogous to the term $\int B_2 \wedge X_8(R)$ in type
IIA \cite{VW} \cite{one}. The nonperturbative result for type IIB
is \cite{GG2} \cite{Seth} \( L=f(\tau,\bar{\tau})\left( I_1 -
\frac{1}{8}I_2 \right) \) where $f(\tau,\bar{\tau})$ is a modular
form in $\tau=C_0+ie^{-\phi}$, and \cite{Roo} \footnote{Here $t_8$
is the usual rank eight tensor that shows up in higher order
corrections, $\epsilon_{10}$ is the antisymmetric constant tensor,
and $R^4$ is a certain quartic polynomial in the curvature
tensor.}

\begin{eqnarray}
I_1&=&t_8 t_8 R^4 + \frac{1}{2} \epsilon_{10}t_8 B_2 R^4
\nonumber\\
I_2&=&-\epsilon_{10} \epsilon_{10}R^4 + 4 \epsilon_{10}t_8 B_2 R^4
\end{eqnarray}
so that the term in $L$ involving $B_2$ cancels out and one is
left with the pure $R^4$ term. There is also a similar
perturbative result at tree level and one-loop. For type IIA there
is another term in $L$ with a $+$ sign between $I_1$ and $I_2$,
which leads to nonzero $B_2R^4$ term \cite{GSIII} (see \cite{Kir}
for details). This is still compatible \cite {Das} with type II
T-duality, because of radius dependence of the corresponding term
in nine dimensions.

\vspace{3mm} The $SL(2,\Bbb{Z})$ symmetry of type IIB string
theory in nine dimensions can be interpreted as a geometric
symmetry of M-theory compactified on a torus $T^2$ \cite{pq}
\cite{Asp} \cite{SchM}. This way there are three scalar fields
corresponding to the moduli of the torus (along directions 9 and
11) given by the volume $V=R_9R_{11}$ and the complex structure
$\omega=\omega_1 + i \omega_2=C_1 + iR_9/R_{11}$, where the metric
on the torus is \(
G_{IJ}=\frac{V}{\omega_2}\left(\begin{array}{cc} |\omega|^2 &
\omega_1
\\
              \omega_1 & 1 \end{array} \right).
\) By T-duality $R_A \leftrightarrow 1/R_B$, $\omega$ is
identified with $\tau$ of type IIB theory, and thus manifests
itself as the S-duality in type IIB $e^{-\phi} \leftrightarrow
e^{\phi}$.

\vspace{3mm} All $R^4$ one-loop terms can be obtained from
one-loop terms in M-theory \cite{Gr}. Such terms contain factors
that are of the form
\begin{eqnarray}
A_{R^4}&\sim&\int dt \sum_{l_1,l_2} \exp(-t G_{IJ}l_I l_J)
\nonumber\\
&\sim&\int dt \sum_{l_1,l_2} \exp \left(- \frac{t}{V} \frac{{|m+n
\omega|^2}}{\omega_2} \right)
\end{eqnarray}

A double Poisson resummation converts the sum over the
Kaluza-Klein charges $(m,n)$ to a sum over the winding modes
$(\hat{m},\hat{n})$ of a worldline along the two cycles of $T^2$,
and the Gaussian integral gives terms proportional to the
nonholomorphic modular form of weight zero \cite{GG2} \cite{Seth}
\cite{Piol} \cite{Berk} (see \cite{Gr} for an overview) \(
\sum_{(\hat{m},\hat{n})\neq (0,0)}
\frac{{\omega_2}^{3/2}}{|\hat{m}+\hat{n}\omega|^3}. \)



\section{Theories in twelve dimensions}

\label{stheor} Let us now begin looking at what sectors of
F-theory are actually known. Quite a lot is already in the
literature. One can get hints from eleven-dimensional M-theory and
ten-dimensional type IIB string theory that there is a theory (or
theories) \footnote{In this paragraph we use the term ``theory''
rather loosely and we do not yet specify the dynamics (nor claim a
full construction of course).} in twelve dimensions that is (are)
playing a role in the topology and the dynamics of those theories.
One can think of two such theories, the manifolds on which they
are defined we take to be $Z^{12}$ and $V^{12}$, respectively.

\vspace{3mm}
 First, there is
the twelve-dimensional coboundary theory that Witten introduced
\cite{flux} to rewrite the Chern-Simons term of M-theory in terms
of the index of the $E_8$-coupled Dirac operator and the index of
the Rarita-Schwinger operator. The topological part of the low
energy limit of M-theory, namely eleven-dimensional supergravity,
is captured by the Chern-Simons term and the one-loop
gravitational correction term, \( \frac{1}{6}\int_{Y^{11}} C_3
\wedge G_4 \wedge G_4 - C_3 \wedge I_8 \label{Mcs} \) where $I_8$
is a polynomial in the curvature of $Y^{11}$ whose class is given
in terms of the Pontrjagin class and the string class as
$[I_8]=\frac{p_2 - \lambda^2}{48}$. The lift of this action to the
twelve-dimensional manifold $Z^{12}$ (where $Y^{11}=\partial
Z^{12}$) is given by \cite{flux}

\( \frac{1}{6} \int_{Z^{12}} G_4 \wedge G_4 \wedge G_4 - G_4
\wedge I_8 \label{ewact} \) by directly using Stokes' theorem.
\footnote{Of course this is not as trivial as it seems because it
requires the vanishing of the relevant spin cobordism groups.
Happily, this satisfied.} A priori this theory has no connection
to type IIB. However, we will show later that there is in fact
such a connection.

\vspace{3mm} Second, there is the ``standard'' F-theory
\cite{Vafa}, which is the lift of type IIB via an elliptic curve.
The complex structure of the elliptic curve is varying over the
type IIB base. In contrast to conventional type IIB
compactifications where $\tau$, as a physical parameter, is taken
to be constant.
   One can relate type IIB on a manifold $X^{10}$ to
F-theory on an elliptically fibered manifold with base $X^{10}$. A
choice of section is usually required \cite{MV} for the
elliptically fibered manifold, i.e. a choice of an embedded base
manifold.

\vspace{3mm} To be compatible with dualities, this theory can also
be considered as the lift of M-theory via a circle. If we choose
to start from M-theory, then we lift the action (\ref{Mcs}) via a
circle $S^1$ to \footnote{Such terms were proposed in \cite{Sag}
in the context of Calabi-Yau compactifications.}

\( \frac{1}{6}\int_{V^{12}} A_4 \wedge G_4 \wedge G_4 - A_4 \wedge
I_8, \label{estandard} \) where $A_4$ is a 4-form potential which
is the lift to twelve dimensions of the 3-form potential $C_3$ of
M-theory in eleven dimensions. On can view $C_3$ in turn as the
contraction of one index of $A_4$, i.e. $C_3=i_* A_4$. This
Lagrangian has been essentially considered in \cite{Pope}, and
global manipulations of this type for circle bundles have been
considered in \cite{HM}.

\vspace{3mm} Looking for field theory in F-theory as a circle
bundle on M-theory leads to certain puzzles and we cannot claim
that all sectors of F-theory arise in this way (accordingly, the
F-theory Lagrangian may need other terms which we do not yet
know). To see this, for example IIA string theory should be a
compactification of M-theory and IIB string theory should be a
compactification of F-theory, but as far as known so far, IIA and
IIB spacetimes can have different homotopy types. This seems
contradicted by proposing a simple relation between M-theory and
F-theory via $S^1$-compactification.

\vspace{3mm} While we do not have a definitive answer to this
problem, there are two ways we can deal with it in the present
paper: first, our main interest is a free field theory based on
elliptic cohomology, which approximates a certain refinement of
the partition function in type II string theory. For this elliptic
field theory to exist, a stronger condition ($w_4=0$) is required
than the known conditions for consistence of type II theories. The
condition we use is the same for type IIA and IIB, so it can be
argued that duality is not violated in our setting (it is at
present unknown if the stronger condition is simply an artefact of
our model, or if it expresses some intrinsic new restriction on
type II strings).

\vspace{3mm} The second possible approach is to deal with IIA and
IIB separately. We shall discuss this in more detail below, and in
fact shall see evidence that different physical signatures may
arise in both cases. In this approach, the Lagrangian
\rref{estandard} is valid for the sector of F-theory which
contains M-theory and type IIA, and the precise Lagrangian for IIB
remains to be determined (however, should be related, since, as we
shall see, the present Lagrangian can be interpreted in a way as
to contain IIB fields). It should be also mentioned that
\cite{Pope} consider certain projections to reconcile on-shell
states between F-theory and M-theory/Type IIB. While we believe
this might be possible in our formulation (see end of section
5.3), we do not attempt a construction, as it seems out of the
scope of the present paper.

\section{Evidence for TMF}
\label{s4}

\subsection{Type IIB and TMF}
\label{stmf}

\vspace{3mm} Let us now consider again the equation \rref{eg3}.
From what we learned in previous investigations, it is likely
correct to say that the field strength $G_3$ should live in a
generalized cohomology theory. For example, when analyzing the IIB
partition function, Witten found that $F_3\in K^1(X)$. $K$-theory,
on its own, of course does not tell a modularity story, and one
needs to solve the puzzle of what happens in the presence of
$H_3$. Some aspects of this were considered in \cite{KSB}. But
from the point of view of \rref{eg3}, it seems that if we want
$G_3$ to live in a generalized cohomology theory, then $\tau H_3$,
$F_3$ must coexist in the same theory. We conjectured in
\cite{KSB} that this theory should be the theory of topological
modular forms, $tmf$, the coefficients of which are, at least
rationally, holomorphic (chiral) modular forms-- see appendix for
a brief review. However, even then, what should one do about the
factor $1/\sqrt{Im\tau}$? This scaling factor is troublesome from
the point of view of algebraic topology, since it is not chiral
and therefore does not occur among the kinds of modular forms
described by $tmf$.

\vspace{3mm} If we are to lift our fluxes to $tmf$, we must
proceed one chirality at a time, and therefore see no choice but
to drop the $1/\sqrt{Im\tau}$ factor. Thus, we consider
\beg{egt3}{\tilde{G}_3=F_3-\tau H_3.} This, of course, now is a
flux with modular weight $-1$, i.e. we have if we denote by
$\tilde{G}_{3}^{\prime}$ the expression obtained by replacing
$\tau$ by \beg{etautr}{\tau^{\prime} =(a\tau+b)/(c\tau +d),}
\beg{egt4}{\tilde{G}_{3}^{\prime} =\tilde{G}_3\cdot
(c\tau+d)^{-1}.} Now this has a striking implication to the
dimension of this class, if it is to be lifted to $tmf$: In that
theory, a class of modular weight $k$ appears in
$tmf^{2k}(X^{10})$. Therefore, our assumptions lead to
\beg{etg5}{\tilde{G}_3\in tmf^{-2}X^{10}.} This points to the
$12$-dimensional picture: suppose, in the simplest possible
scenario following Vafa \cite{Vafa} that \( V^{12}=X^{10}\times E
\label{evafa1} \) where $E$ is an elliptic curve. Then let \(
\mu\in tmf^2(E) \) be the generator (given by orientation). This
then suggests introducting $\mu$, instead of $1/\sqrt{Im\tau}$, as
the correct scaling factor of $G_3$, and passing to $12$
dimensions: we have \beg{etg50}{\tilde{G}_3 \times \mu \in
tmf^{0}(V^{12}).} It is a surprise that the class ends up in
dimension $0$ and no odd number shows up here. However, note that
the fiber $E$ contains odd-degree non-torsion cohomological
classes, so all kinds of shifts between even and odd are possible
here. Modular classes of weight $0$, however, must be in dimension
$0$.

\vspace{3mm} We used here the statement that for a space $X$,
classes in $tmf^{k}(X)$ are modular of weight $k/2$, which means
that upon the transformation \rref{etautr}, the class transforms
by introducing the factor $(c\tau+d)^{k/2}$. It is fair to point
out that to make this rigorous mathematically, some discussion is
needed. In fact, we will find it necessary to generalize to an
elliptic cohomology theory $E$ which is in general modular only
with respect to some subgroup $\Gamma\subset SL(2,\Z)$ (see
below). So, we give the discussion in this context. The $tmf$
discussion is analogous. The first question we must ask is what is
$\tau$ mathematically? The answer is that $\tau$ appears only when
we apply the forgetful map \footnote{This is formal expansion of
K-theory in the the power $q^{1/24}$ of the formal parameter $q$.
The second set of paranthesis indicates that the generator
$q^{1/24}$ is inverted. Such expansions relating elliptic
cohomology to K-theory were used in \cite{KS} to interpret the
elliptic refinement of the type IIA partition function.}

\beg{eforget}{E^k(X)\r K^k(X)[[q^{1/24}]][q^{-1/24}].} Then one
takes $q=exp(2\pi i \tau)$. The right hand side denotes power
series in $K$-groups, with the parameter $q^{1/24}$ inverted. This
map was discussed in our previous papers \cite{KS,KSB}. On this
level of coefficients, it is given simply by the fact that a
modular form may be expanded in the modular parameter $\tau$. For
forms which are modular only with respect to a subgroup of the
modular group, fractional powers of $q$ are needed: in the case of
complex-oriented cohomology, one encounters $q^{1/24}$.

But in addition to this, \rref{eforget} must be suitably
normalized. As explained in \cite{KS}, one has a canonical map of
generalized cohomology theories \beg{eforget1}{E\r
K[[q^{1/24}]][q^{-1/24}]} whose induced map on coefficients
(homotopy groups) makes the $k$-th homotopy group modular of
weight $k/2$. This is not the correct normalization to use in
\rref{eforget}, because then $\tilde{E}^0(S^k)=E^{-k}(*)$ would
have modular weight $-k/2$, whereas we would like $0$. To this
end, we need to compose with some map which would multiply by some
normalizing factor of weight $k/2$ in the $k$-th homotopy group.
Such operation indeed exists, and it is the Adams operation
\footnote{ In fact, a more precise discussion uses Ando power
operations in elliptic cohomology, but we will not need that
here.}

\beg{eadams}{\psi^{\eta}:K[[q^{1/24}]][q^{-1/24}]\r
K[[q^{1/24}]][q^{-1/24}].} Here $\eta$ is the Dedekind function
($\Delta^{1/24}$ where $\Delta$ is the discriminant form), which,
note, is a unit in $K[[q^{1/24}]][q^{-1/24}]$. Now we see that
composing \rref{eforget1} with \rref{eadams} gives the correct
normalization of \rref{eforget} for $k=0$. For general $k$, if we
simply delooped this map, we would be be in weight $0$ instead of
$k/2$, so we need to multiply the delooped map by $\eta^k$ to get
the correct normalization.

\vspace{3mm} To summarize the results of this section, our
conclusion confirms that if we want to seriously consider the
modularity of the flux $G_3$ in $tmf$, the correct way is to
introduce the normalization \rref{etg50}, and work in F-theory. We
will see in the later sections that the picture described above
may be overambitious: we do not know of a sector of F-theory which
would really use $tmf$ this way, and which would explain
modularity of IIB with respect to the whole group. Nevertheless,
we will see that the naive discussion given in this section is
roughly correct.

\subsection{Twelve dimensions and TMF}

Let us dedicate one section to speculation about an F-theory which
would be governed by the ideally modular elliptic cohomology
theory $tmf$. As already remarked, we will see later that we will
fall somewhat short of this goal, and will have to revert to less
ideal elliptic cohomology theories and elliptic curves. Perhaps
the ``ideal theory'' could be reached by some type of advanced
orbifolding of the fiber $E$ in \rref{evafa1}, just as $tmf$ in
mathematics is constructed that way from elliptic cohomology. For
now, however, let us make a few first observations about the field
content of such ideal F-theory.

\vspace{3mm} At least when tensored with $\Q$ (or more generally a
field of characteristic $0$), the coefficients (homotopy classes)
of $tmf$ are modular forms:
\beg{etmf1}{tmf_{*}\otimes\Q=\Q[g_2,g_3].} Here recall that $g_2$,
$g_3$ are the standard modular forms of weights $4$ and $6$, given
by the Eisenstein series as \bea g_2(z)=\frac{4}{3}\pi^4 E_4(z)
\\
g_3(z)=\frac{8}{27}\pi^6 E_6(z) \eea with the (normalized)
Eisenstein series given by \( E_k(z)=\frac{1}{2} \sum_{m,n \in \Z;
(m,n)=1} \frac{1}{(mz+n)^k} \) where $(m,n)$ denotes the greatest
common divisor.

\vspace{3mm} In particular, the notation has nothing to do with
our previous notation for fluxes. Now as remarked above, in $tmf$,
the dimension of a class is twice its modular weight, so
rationally, \bea g_2\in tmf_8,
\nonumber\\
g_3\in tmf_{12}. \eea Now the F-theory we are considering takes
place on a $tmf$-orientable manifold $Z^{12}$, and the topological
fluxes we consider are in its $tmf$-cohomology. Recall (cf.
\cite{KSB}) that the obstruction to $tmf$-orientability is \(
\lambda\in H^4(Z, \Z)\mod 24. \) In any case, orientability
implies that we have a class \( u\in tmf^{12}(Z), \) so using
\rref{etmf1}, and the dimensions, we see that in non-negative
dimensions, we have possible field strength sources $u$, $ug_2$,
$ug_3$ in dimensions $12$, $4$, $0$ (the dimension of the
coefficients is subtracted from the dimension of a class in
generalized cohomology). Note that this derivation is of course
quite schematic, but on the other hand somewhat analogous to the
derivation of the dimension of RR-sources in type II string theory
from $K$-theory. Also, we have only considered $tmf$ rationally.
Delicate questions regarding the integrality of the proposed
fields would have to be considered, specifically at the primes 2
and 3.

\vspace{3mm} If we accept this, then we see there is a fundamental
field strength in dimension $4$. It is, of course, natural to
conjecture that this is related to the field strength $G_4$ in the
M-theory compactification of the appropriate sector of F-theory.
We look at this next.

\subsection{Anomalies in type IIB and congruences}

In principle there can be anomalies associated with the $U(1)$
symmetry and with the $SL(2,\Bbb{Z})$. The $U(1)$ anomaly
\cite{Mar2} can be cancelled by adding the term \cite{GG} \(
S=\frac{1}{4 \pi} \int_{X^{10}} \phi F_2 \wedge I_8(R) \) provided
that \( \frac{1}{4 \pi} \int_{X^{10}} F_2 \wedge I_8(R) \in
\Bbb{Z} \) since $\phi$ is $2\pi$-periodic. Here $F_2$ is the
curvature of the upper half plane, given in terms of the modular
parameter by \( F_2=\frac{i d{\bar \tau} \wedge d\tau}{4
(Im\tau)^2}, \) and $I_8(R)$ is the Green-Schwarz anomaly
polynomial in $R$, the curvature of $TX^{10}$.

\vspace{3mm} The $SL(2,\Bbb{Z})$ anomaly is cancelled by adding
the term \cite{GG} \( S''=\frac{i}{4 \pi} \int_{X^{10}} \ln
g(\tau) F_2 \wedge I_8(R) \) where $g(\tau)$ is a modular form
that satisfies (up to a constant phase) \( g(\Lambda\tau)=\left(
\frac{c\tau + d}{c {\bar \tau} +d} \right)^{1/2} g(\tau) \) where
$\Lambda$ is the $SL(2,\Bbb{Z})$ mobius action. The
$SL(2,\Bbb{Z})$ symmetry is unbroken if \cite{GG} \( \frac{1}{4
\pi } \int_{X^{10}} F_2 \wedge I_8(R) \in N \Bbb{Z} \) where $N$
is $4$ or $12$ depending on the transformation property of
$g(\tau)$. Therefore, we see that if we take the latter case then
the integral \( \frac{1}{2 \pi } \int_{X^{10}} F_2 \wedge I_8(R)
\) is in $24 \Bbb{Z}$. If $I_8(R)$ is integral, then the $U(1)$
curvature $F_2$ is in $24 \Bbb{Z}$. We again see the mod $24$
congruence.

\subsection{The M-theory topological Lagrangian}
\label{smass}

In \cite{flux}, Witten derives the effective Lagrangian of
M-theory which comes from the Chern-Simons term. Simply to get
consistency, i.e. to make the Lagrangian well-defined, one gets
the action (in our notation) \rref{ewact}
where $Z^{12}$ is a $Spin$-manifold whose boundary is M-theory
spacetime $Y^{11}$. In this section, we shall try to understand
this Lagrangian in the context of the kind of theories we are
considering in this paper.

\vspace{3mm} One has, (at least as differential forms),
\beg{em2}{d*G_4= -\frac{1}{2} G_4 \wedge G_4. } It is therefore
appealing to write the Chern-Simons Lagrangian term (``on-shell'')
as \beg{em1}{\frac{1}{12}G_4\wedge(*G_4),} which looks rather like
a gauge-theoretical kinetic term. However note that this still
does not explain the consistency of such expression.

\vspace{3mm} But this is related to the mathematical notion of
Massey products. A differential graded algebra (DGA) is a (not a
priori commutative) graded algebra $A$ with a map $d:A\r A$ of
degree $+1$ which satisfies the relations \bea
        dd&=&0,
\\
       d(ab)&=& (da)b +(-1)^{dim a}a(db).
\eea (Different sign conventions are possible.) Then the
cohomology $H(A)$ of $A$ with respect to $d$ is a graded algebra.
It has further certain operations called (matrix) Massey products.
These are essentially the only operations, but if $A$ has any kind
of commutativity property, more operations arise, although many of
them are torsion. In any case, the simplest Massey product is a
correspondence \(
      H(A)\otimes H(A)\otimes H(A) \r H(A)
\) which is denoted by $[a,b,c]$, where $a,b,c\in H(A)$. It is
defined only when $ab=bc=0\in H(A)$, and the dimension of the
result is \(
      dim(a)+dim(b)+dim(c)-1.
\) It is also not well defined, it is only defined modulo terms of
the form $ax +yb$ where $x,y$ are some elements of $H(A)$ over
which we have no control. They may, however, sometimes be
excluded, for example for reasons of dimension.

\vspace{3mm} The definition of $[a,b,c]$ is as follows: we have \(
     ab=dy, bc=dz \;\text{for} \; y,z\in A.
\) Then set \( [a,b,c]= yc + (-1)^{dim a + 1}az. \) It is obvious
that this is a cocycle, and that the cohomology class is defined
modulo the indeterminacy given above. It is worth noting that all
Massey products are essentially elaborations of this principle. A
Massey product $[a_1,....,a_n]$ exists if and only if all
``lower'' Massey products of these elements vanish, and also one
may do the same thing for matrices of elements. That is the whole
story for DGA's.

\vspace{3mm} In our situation, the equation \rref{em2} implies
\beg{em3}{-\frac{1}{2}[G_4,G_4,G_4]=[G_4, *G_4]} (the right hand
side has the Lie bracket, the left hand side the Massey product).
This suggests rewriting \rref{ewact} as
\beg{em4}{\frac{1}{6}[G_4,G_4,G_4],} which now is at least an
expression which lives entirely in cohomology. However, let us
take this one step further and see what are the implications of
this in F-theory.

\vspace{3mm} In \cite{flux}, as noted above, $Y^{11}$ is the
boundary of a manifold, a `Spin cobordism', $Z^{12}$. To prove
invariance, one also considers the case when $Z^{12}$ is obtained
from gluing two cobordisms together, i.e. $Z^{12}$ is an
orientable compact manifold and $Y^{11}$ is a submanifold of
codimension 1 such that $Z-Y$ has two connected components (each
of which is a cobordism). Then from the Mayer-Vietoris sequence,
there is a connecting map \(
    T: H^k(Y^{11}) \r H^{k+1}(Z^{12})
\) (which can be thought of as a kind of transfer). Now let
$a,b,c\in H^*(Z^{12})$ (we should think $a=b=c=G_4$). Suppose
further $a^{\prime}b^{\prime} =b^{\prime}c^{\prime}=0\in H^*(Y)$
(the ${}^{\prime}$ means restriction from $H^*Z^{12}$ to
$H^*Y^{11}$). Then we have \beg{emtt}{
T[a^{\prime},b^{\prime},c^{\prime}]= abc \mod
\text{indeterminacy}} where the Massey product is taken in $H^*Y$,
the product in $H^*Z$. The indeterminacy can be taken as $az+xc$
where $z,x$ are cocycles in the opposite connected components of
$Z^{12}-Y^{11}$.

\vspace{3mm} A sketch of a proof can be obtained as follows: let
us think of the Poincar\'e dual chains. Make the cycles
representing $a,b,c$ in $Z^{12}$ intersect transversally with
$Y^{11}$. Now restrict the chains $a,b,c$ to chains (not cycles)
$a_i, b_i, c_i$ on the closures $Z_i$ of connected components of
$Z-Y, i=1,2$. Then $d(a_1b_1) = a^{\prime}b^{\prime}$,
$d(b_2c_2)=b^{\prime}c^{\prime}$. Furthermore, in $Z_1$, the
intersection of $a_1b_1$ with $c_1$ is the same as the restriction
in $Z_1$ of $u$ with $c_1$ where $du=a^{\prime}b^{\prime}$ in
$Y^{11}$, which in turn is the same as the intersection of u with
$c^{\prime}$ in $Y$. Similarly on $Z_2$. Now on chains, $T$ is
represented by inclusion. So \(
       a_1b_1c_1+a_2b_2c_2
\) represents $T[a^{\prime},b^{\prime},c^{\prime}]$, as claimed.

\vspace{3mm} This suggests again that the effective
$12$-dimensional topological Lagrangian term should be \beg{em6}{
\int_{Z^{12}} \frac{1}{6}G_4\wedge G_4\wedge G_4 } with the
indeterminacy described above, which of course coincides with the
result of \cite{flux}. But more interestingly, the $1$-loop
correction term in \rref{ewact} can be explained as a part of the
indeterminacy of \rref{emtt}. Thus, it is interesting to note that
indeed \rref{em4} is a correct way to rewrite \rref{ewact}, and
that the $1$-loop correction terms in M-theory is a part of Massey
product indeterminacy.

\vspace{3mm} Another comment is perhaps in order. It could be
argued that a defect of the Massey product approach is that it
does not predict the $1$-loop gravitational term as the correction
term. This is a delicate issue and we would say this criticism is
partially true: on the one hand, certainly the Massey product
approach does not, without further rigidification of the input,
predict the precise form of the counterterm. On the other hand, it
does predict that such a term must exist. \footnote{While there is
a version of the Massey product which does not use indeterminacy,
it requires more input data, and at present we do not know if it
helps predict the one-loop term more accurately.} (A caveat is the
coefficient $1/6$, which cannot be predicted by rational
cohomology; a proper integral refinement, possibly using
generalized cohomology, would be needed. Recall that the arguments
applied in \cite{DMW} are rather delicate. Although generalized
cohomology is the main theme of this paper, and this is perhaps
one of the fundamental issues of M-theory, we do not have this
precisely worked yet.)

\vspace{3mm} Accepting, however, that the Massey product does
predict the existence of a counterterm, it is then actually not
bad at predicting the term itself. The indeterminacy is
\beg{eindet}{G_4\wedge I_7 } where $I_7$ is a $7$-dimensional
cohomology class in $Y^{11}$ (which must be distinguished from the
$8$-dimensional cohomology class $I_8$ in $Z^{12}$).\footnote{One
might be tempted to say that $dI_7=I_8$ via Stokes theorem.
However, this is not correct for two reasons. First, $I_8$ is not
a coboundary (although of course it is locally), and second, $I_7$
is {\it closed}, i.e. it is a cohomology class in $H^7(Y^{11})$.
The situation is quite analogous to that of the potential/field
strength:  $I_8$ plays the role of the field strength, $I_7$ is
the indeterminacy of the potential, which is a closed gauge term,
i.e. a shift gauge transformation that can be added  to it.} The
term \rref{eindet} does not appear to be excluded by the $1$-loop
approach. Although we do not know its exact meaning, it is
probably related to the dynamics of M$5$-branes, as is the
$1$-loop term. In fact, \rref{eindet} looks like a coupling of
M$2$-brane and closed M$5$-brane field strengths.

\vspace{3mm} One reason for discussing these manipulations here is
that it is possible a similar device could be used to unify the
two seemingly different F-theoy Lagrangians \rref{ewact},
\rref{estandard} in Section \ref{stheor}. If we denote by $G_5$
the field strength corresponding to the potential $A_4$, the
suggested F-theory topological term is
\beg{eg5m}{\frac{1}{6}\int_{V^{12}}[G_4,G_4,G_5].} As written, the
Massey product takes place in the algebra of differential forms.
\footnote{Of course, a refinement of $G_5$ in elliptic cohomology
of the $12$-dimensional spacetime would be desirable.
Schematically, this seems consistent since the dimension of the
class would increase by 1 by wrapping around the additional degree
of freedom. However, one must be careful while considering the
exact nature of this additional dimension. We will return to this
point later.} This of course needs further discussion, but the
point is both a 1-loop gravitational correction term and a term of
the form \rref{ewact} can be considered indeterminacy terms to
\rref{eg5m}. In the case of the 1-loop term, the discussion is
similar as the case of M-theory earlier in this section. In the
case of \rref{ewact}, this phase vanishes on a closed manifold
$Z^{12}=V^{12}$. This indeed corresponds to adding the cocycle
$G_4$ to the potential $A_4$, which is a gauge transformation not
affecting the field strength.

\section{The partition function of F-theory compactified on an elliptic
curve}

\label{s5}

\subsection{Elliptic cohomology}

Let us now approach the problem from another angle. Namely, let us
go back to $10$-dimensional type II string theory. In \cite{KS},
\cite{KSB}, we have observed that the partition functions of IIA
and IIB string theories (see \cite{DMW}, \cite{MW}) can be lifted
to elliptic cohomology. We constructed this lifting by carefully
observing the homotopical content of the IIA partition function
obstruction of \cite{DMW}. However, what is the correct
interpretation of these partition functions?

\vspace{3mm} In this section, we propose an answer to this
question: the elliptic partition function belongs to F-theory
compactified on an elliptic curve, which unifies both IIA and IIB
string theory. Roughly, the idea is this: In elliptic cohomology,
we see another parameter in the coefficients of the theory. In
\cite{KS}, we worked mostly with the cohomology theory $E(2)$, in
which case the extra parameter will be $(v_1)^3(v_2)^{-1}$, where
$v_1$ is the Bott generator and $v_2$ is the degree six analog.
One can work with other elliptic spectra and get different
parameters. But the point is that in all cases, the additional
parameter is some modular form of some level, i.e. a power series
in $q=e^{2\pi i\tau}$ where $\tau$ is the modular parameter of an
elliptic curve. So one can ask what causes a theta function (more
precisely theta constant) of a lattice $\Gamma$ to be modified in
this fashion, i.e. where the value is, instead of a number, a
function in a modular parameter $\tau$ of an additional elliptic
curve? The answer is that this arises precisely when we tensor
$\Gamma$ by another lattice of dimension $2$ whose period is
$\tau$. Tensoring with a two dimensional lattice amounts to
summing two copies of $\Gamma$ (at least as abelian groups). One
can argue that if $q\r 0$, then $\tau \r i\infty$, so the other
copy of $\Gamma$ is ``infinitely far'', thus reducing the new
function to the old one in the $q\r 0$ limit.

\vspace{3mm} But where does the new lattice come from? It comes
from the $1$st cohomology of an elliptic curve, which is the
theory $E$ on which we are compactifying $F$-theory. In other
words, in $F$-theory on $V^{12}=X^{10} \times E$ which contains
type IIB string theory on $X^{10}$, the odd degree field strengths
move to even-dimensional cohomology of $V^{12}$, as predicted
above in Section \ref{stmf}. For example, from the F-theory term
\( \int_{V^{12}} A_4 \wedge G_4 \wedge G_4 \) which was proposed
in \cite{Sag} and used in \cite{Pope}, we obtain the type IIB
Chern-Simons term \( \int_{X^{10}} A_4 \wedge F_3 \wedge H_3 \)
after reducing on the elliptic curve one step at a time to get
$H_3$ and $F_3$ as results of contraction of one index of $G_4$,
and $A_4$ remains the same. So $G_5$ is lifted by $H^0(E)$ and
$G_3$ by $H^1(E)$. What about $G_1$? Note that the reason to
consider F-theory in the first place was to try to interpret $G_1$
(i.e. the axion-dilaton combination) as the (non-constant) moduli
of the elliptic curve. Thus we propose that $G_1$ is not lifted to
F-theory but only shows up in ten dimensions upon compactifying on
a nontrivial torus. This is compatible with \cite{Pope} who
consider a field content in twelve dimensions consisting of a
metric, a dilaton, a four-form and a five-form field strengths,
but no p-form field strengths ($p=2,3$) which would come from
lifting $G_1$ via $H^1$ and $H^2$. Of course, there could be a
nontrivial mixing between the dilaton in twelve dimensions and the
dilaton coming from the moduli of the torus. This might not be
surprising from a Kaluza-Klein point of view, but we do not
explore it further as it would be outside the scope of the paper.

\vspace{3mm} In IIA, this might seem more confusing, since we have
a field $G_4$ in dimension $4$ and $H_3$ in dimension $3$.
However, we think the answer has to be as follows. Once again, we
should have compactified F-theory on $V^{12}=X^{10}\times E$. But
this time, consider an intermediate step, M-theory compactified on
$X^{10}\times S^1$. In this compactified M-theory, the $H_3$ picks
up a dimension by multiplying with a first cohomology class of
$S^1$ and is absorbed into $G_4$. In the F-theory considered here
(the standard F-theory), indeed $G_4$ expands into $G_5$ by
coupling with $H^1(E)$. On the other hand, the IIA-theoretical
$H_3$ becomes absorbed in this $G_5$ also by coupling with
$H^2(E)$. Thus, we see the same modularity (see below for more
notes on modularity) as in the standard F-theory related to IIB,
and since that theory has both $G_4$ and $G_5$, this further
supports the idea that this theory be a unification between type
IIA and IIB string theory (see \cite{Vafa} \cite{Sag}
\cite{Pope}). However, we note from section \ref{stheor} above
that this sector is not obtained as cobordism of $Y^{11}$, but as
$Y^{11}\times S^1$.

\subsection{E-theoretic formula for the fields and new characteristic
classes}

According to \cite{DMW}, formula (7.2) states that the total field
strength $G(x)$ of type II 10-dimensional string theory is $2\pi$
times \beg{edmw72}{\sqrt{\hat{A}(X)}ch(x).} This formula is
needed, since the metric of the $K$-theory lattice is, up to a
factor of $1/(2\pi)^2$, given by \beg{edmw73}{\int_X G(x)\wedge *
G(y).} However, formula \rref{edmw72} applies to a $K$-theory
setting, so it needs adjustment in case of elliptic cohomology.
There is no problem with the Chern character, since for any
elliptic cohomology theory $E$, there is a canonical map $E\r
K((q))$ (where $q$ is as above), so we may compose with the Chern
character to get a map \beg{eche}{ch_E:E\r H^*((q)).} On the other
hand, the term $\sqrt{\hat{A}(X)}$ should be replaced by an
analogous term related to the Witten genus, which is \(
\sigma(X)^{1/2} \) where $\sigma(X)$ is the characteristic class
of $X$ associated with the power series \(
\sigma(z)=(e^{z/2}-e^{-z/2})\cform{\prod}{n\geq
1}{}\frac{(1-q^ne^z) (1-q^ne^{-z})}{(1-q^n)^2}. \) Therefore, our
formula for the elliptic field strength associated with $x$ is
\footnote{We have demonstrated this formula only up to terms that
vanish as $q\r 0$, so in principle such terms could be present.
However, it is not obvious that there are natural such
candidates.} \beg{eell1}{G(x)=\sigma(X)^{1/2}ch_E(X). } Note that
this $\sigma$-function, in the $q\r 0$ limit, reduces to the
characteristic function of the $\hat{A}$-genus, thus reducing this
field strength to the type II field strength in the
$10$-dimensional limit.

\vspace{3mm} We should of course remark that the partition
function we consider is, similarly as in \cite{DMW}, approximate
in that we work in the free field limit. This means that the
action we consider is essentially just the Hermitian metric on the
field strengths. Using the standard definition of partition
function, we therefore obtain the theta function.

\vspace{3mm} The definition of the elliptic partition functions
given in \cite{KS}, \cite{KSB} are then complete. As mentioned
above, we propose that these functions are, in fact, partition
functions of the F-theory sectors on $X^{10}\times E$ which, when
$E$ goes to $0$, reduce to type IIA and IIB $10$-dimensional
string theories.

\subsection{Interpretation of Twist and modularity}


\vspace{3mm} Let us recall now again the IIB string theory
modularity puzzle (see section \ref{stmf}). In type IIB string
theory, we have an RR-field strength $F_3$ and an NSNS-field
strength $H_3$ which are in a relation of modularity. As pointed
out in \cite{DMW}, the $K$-theory based partition function for
type IIB does not explain that modularity, and it cannot be
explained by twisted $K$-theory either, as shown in \cite{KSB}. Of
course, as also mentioned there, the possibility is not excluded
that by introducing more terms, such as the $P$-term that depends
only on the topology and the spin structure of the manifold
\cite{DMW} \cite{DFM}, into the modularity equation, one could
start building by hand a Postnikov tower of a different
classifying space or generalized cohomology which could give the
correct explanation.

\vspace{3mm} This is however not the approach we take here.
Instead, we build directly a theory (at least its free field
approximation) based on elliptic cohomology of the
$10$-dimensional spacetime $X^{10}$. What we conjecture (see also
\cite{KSB}) is that this theory is related to F-theory
compactified on an elliptic curve $E$. It is rather natural then
to conjecture that modularity in the first cohomology of $E$
explains the modularity in type IIB theory. (We also commented
briefly above on why this modularity is broken in IIA.) This
construction does not come for free. In order for the lift to
F-theory to be consistent, we get an obstruction \( w_4=0 \) which
seems foreign to type II string theory (although it occurs in
heterotic string theory, thus perhaps hinting that F-theory
provides an even further unification). Also, the combined field
strength $G_3$ (see section \ref{stmf} above for more discussion)
must be lifted to elliptic cohomology, which restricts the kind of
configurations allowed. Twisting in the new theory disappears. The
combined $G_3$ field strength is (an additive) generalized
cohomology class, whereas twisting allows, at least a priori,
non-additive configurations.

\vspace{3mm} After introducing all this, we got modularity which
is indeed tied to the modularity in the first cohomology of the
F-theoretical fiber $E$. However, note that even then the picture
we get is not quite as ideal as one might hope. Mathematically,
the problem is that elliptic cohomology spectra are not completely
modular with respect to the whole group $SL(2,\Z)$. Only the
spectrum $TMF$ enjoys such full modularity, but that is not an
elliptic spectrum. In fact, if we agree to specialize to
information at $p=2$ ($2$-torsion does seem like the most
interesting information), then following Hopkins and Mahowald
\cite{HMA}, we may use the elliptic spectrum $E_2$ with
coefficients $W_2[[a]][u,u^{-1}]$. Then $TMF$ (at $p=2$) can be
obtained as homotopy fixed points of $E_2$ with respect to an
action of the group $SL(2,\Z/3)$. In other words, we may roughly
say that $E_2$ is modular with respect to the congruence group
$\Gamma(3)$, i.e. that we are only allowed to perform modular
transformations which fix the group of points of order $3$ on the
elliptic curve. Accordingly, we only recover level $3$ modularity
of the combined $G_3$ field strength of type IIB string theory.

\vspace{3mm} One could conjecture that an ideal F-theory (as was
suggested above) could be obtained by an orbifolding analogous to
the construction in homotopy theory which produces the spectrum
$TMF$ (or its connected form, $tmf$). Let us try to work out the
implications of such construction. First of all, mathematically,
we have the advantage that we have a toy model. To simplify the
discussion, let us look again at generalized cohomology theories
completed at $p=2$, as in the last paragraph. Then we saw we get
$TMF$ from $E_2$ by taking homotopy fixed points with respect to
the group $SL(2,\Z/3)$. However, that group has a normal subgroup,
namely the center, which is isomorphic to $\Z/2$. The non-zero
element $\alpha$ of this center is the diagonal matrix with
entries equal to $-1$. We can therefore obtain $TMF$ in two
stages, first taking homotopy fixed points
\beg{ehom1}{(E_2)^{h\Z/2},} and then again homotopy fixed points
of the generalized cohomology theory (=spectrum) \rref{ehom1} with
respect to $PSL(2,\Z/3)$. However, as noted above, the map
$\alpha$ is the inverse operator on the elliptic curve (in
homotopy theory, one sees a so called supersingular elliptic curve
over $\mathbb{F}_4$, and in fact all its information is extracted
from its formal group law, which is of height $2$; see \cite{KS}
for a review of formal group laws in the physical context. The
element $\alpha$ is then the inverse series of that formal group
law). The point of discussing this in such detail is that taking
fixed points with respect to the inverse series of a formal group
law is a well known operation in homotopy theory: one obtains the
real form of the theory. For example, starting with $K$-theory,
one obtains $KO$. Starting with $E_2$, the theory \rref{ehom1}
becomes in fact the real elliptic cohomology theory \(
(E\R_2)^{\Z/2} \) discussed in \cite{hk} (as shown there, there is
a ``completion theorem'' which makes it unnecessary in this case
to distinguish between actual and homotopy fixed points).

\vspace{3mm} The appearance of the real form of a generalized
cohomology theory is interesting here. In the case of $K$-theory,
its real form $KO$ describes the sources of type I string theory,
which can be obtained from type IIB string theory by orbifolding.
Note however that this is worldsheet orbifolding, using the
automorphism of the theory which exchanges the chiral sector, i.e.
a worldsheet involution that reverses the signs of the worldsheet
coordinates, and thus interchanging left movers with right movers.
It does not seem from this worldsheet point of view that in $10$
dimensions, one could consistently orbifold any further. Another
way of expressing this is to say that supersymmetry cannot be
broken further than $N=1$, starting from $N=2$.

\vspace{3mm} From the $12$-dimensional point of view, when
constructing \rref{ehom1}, however, we see another side of the
story. We can, in fact, identify physically what kind of
orbifolding \rref{ehom1} corresponds to. This is because we know
that the element $\alpha$ is the inverse of the elliptic curve,
and that elliptic curve we understand (from modularity) to be a
form of the fiber $E$ of \rref{evafa1}. Therefore, we are
orbifolding with respect to the involution of the two fiber
dimensions in spacetime! In this context, the additional
orbifolding with respect to $PSL(2,\Z/3)$ could possibly be
consistent, although details would certainly have to be worked
out. But how is it possible that worldsheet orbifolding of type
IIB in $10$ dimensions could correspond to the spacetime fiber
orbifolding in dimension $12$?

\vspace{3mm} While we do not have a complete explanation (it is
perhaps a ``string miracle"), we can point out that this
phenomenon, at least in $11$ dimensions, has in some sense already
been observed. Compactified M-theory on $S^1$ is on the
strong/weak duality line between type IIA string theory and
M-theory. Applying spacetime orbifolding to the eleventh dimension
with respect to the inverse operator gives Ho\v{r}ava-Witten
M-theory, which is S-dual to $E_8\times E_8$ heterotic string
theory. Applying T-duality, we get $Spin(32)/{\Z_2}$ heterotic
string theory, which is S-dual to type I string theory. The latter
is obtained by worldsheet orbifolding of type IIB string theory
\cite{Sag2} via projecting by an involution (i.e. orientifold)
$\Omega$ that exchanges the left and the right closed string
oscillators and acts on the open string oscillators by introducing
a $\Z_2$ phase.\footnote{More precisely, $\Omega$ acts on the
closed sector by exchanging $\alpha_m^{\mu}$ and ${\tilde
\alpha}_m^{\mu}$ and on the open sector by exchanging
$\alpha_m^{\mu}$ and $\pm (-1)^{m} \alpha_m^{\mu}$.} We propose
that type I can be lifted to a theory $\widetilde{\rm M}$ which is
T-dual to the original M-theory compactified on $S^1$ (as remarked
below, we do not know if $\widetilde{\rm M}=M$). In any case, if
we suppress U-dualities from the notation, we get, schematically,
the following diagram:
$$
\diagram
*\rrto^{sO}\dto_{T}&&*\dto^{T}\\
*\rrto^{wO}&&*
\enddiagram
$$
where $sO$ stands for spacetime orbifolding in the eleventh
dimension, and $wO$ stands for worldsheet orbifolding in $10$
dimensions, and $T$ stands for T-duality. This is the kind of
relation between worldsheet and spacetime orbifolding proposed
above.

\vspace{3mm} Note that the spacetime involution applies to both
dimensions of $E$, so it preserves orientation, while the
worldsheet involution only applies to one coordinate, i.e. it
reverses orientation.

\vspace{3mm} One might also justify the truncations done in
\cite{Pope} for the reduction from F-theory to M-theory and type
IIB string theory. There, (consistent) truncations were imposed by
hand on the fields, which amounted to setting $G_5$ to zero in
compactifying to M-theory on $S^1$ and setting $G_4$ to zero in
compactifying F-theory to type IIB string theory on an elliptic
curve. We propose that such truncations can be made natural by
looking at a Ho\v{r}ava-Witten-like construction, but for the
elliptic curves instead of the circle. More precisely, we propose
the existence of involutions on both elliptic curves, the one
fibered over IIA and the one fibered over IIB, in such a way that
orientation-reversing kills $G_4$ in the case of IIB and kills
$G_5$ in the case of IIA. From the point of view of M-theory, this
means that the extra twelve-dimensional circle comes with an
involution on it. \footnote{We do not imply that all the theories
we discuss are related either by taking boundaries or simply by
using $S^1$ factors. The discussion is schematic and a more
precise description will involve other manifolds such as $K3$ in
order for the picture to be compatible with the web of dualities.
In some limit, $K3$ can be viewed as some orbifold of $T^4$ so
what we mean by a circle with an involution is something crudely
similar within the Calabi-Yau manifolds. } We do not attempt a
construction here as this would be beyond the scope of this paper.

\vspace{3mm} Let us make one more comment, which is more related
to the IIA sector. Diaconescu, Freed and Moore \cite{DFM} consider
a cubic refinement of the triple pairing in $G_4$ associated with
M-theory. This in our language is related to the cubic structure
\cite{AHS} on elliptic cohomology in the same way as the quadratic
refinement of the pairing $\omega$ in \cite{DMW} is related to the
quadratic structure on $K$-theory corresponding to $KO$.

\vspace{3mm}

\subsection{The Witten genus and a possible explanation via loop
groups}

There is another provocative coincidence which may support our
explanation of the elliptic partition function as compactification
of the standard F-theory \rref{estandard} on an elliptic curve.
When considering the elliptic field strength \rref{eell1}, we see
that the free action (Hermitian metric) in that theory relates to
the Witten genus in the same way as the action of the
$K$-theoretical field strength \cite{DMW} relates to the
$\hat{A}$-genus.

\vspace{3mm} But when Witten first introduced his genus
\cite{W-loop}, he made another suggestion of relation with the
$\hat{A}$-genus, namely that his genus should be related to taking
index of loop bundles on loop space. This, in fact, has led to
much speculation on the nature of elliptic cohomology, which is
well summarized in \cite{sell} (see the volume \cite{LNM} for the
orginal references). Most of this speculation, which continued to
the present day (cf. \cite{Stolz}), was in the worldsheet
modularity direction, but when trying to match this with evidence
from loop groups, \cite{ell} found that the elliptic curve shows
up in spacetime as well. Here we shall propose that the elliptic
curve in spacetime should, in fact, be the fiber of standard
F-theory compactified on the elliptic curve. In fact, strikingly,
\cite{ell} found defects to modularity very similar to those found
in the present paper.

\vspace{3mm} What we propose is the following. When forming the
compactification of F-theory on an elliptic curve, there is an
intermediate step: compactification of F-theory on a circle, which
should be M-theory following \cite{Pope}. In view of our previous
discussion in Section \ref{stheor}, it is safest here to consider
this as a sector of F-theory which contains type IIA string
theory; the sector containing IIB-theory may possibly be
different. In fact, in the IIB case, one should also have an
$S^1$-reduction of F-theory, and one can have a symmetric picture
between type IIA and type IIB string theory in connection to
F-theory. Another way to pose this question is whether it makes
sense to ask for a ``T-dual'' of M-theory. We do not know if
M-theory would be ``T-dual'' to itself, although this seems to be
hinted at in \cite{Vafa}. We sometimes use the term
${\widetilde{\rm M}}$-theory to refer to the ``T-dual'' of
M-theory which contains IIB. We do not know if $\tilde{M}$-theory
is the same as M-theory under suitable conditions. However, the
existence of such a theory would not follow from a strong coupling
argument, since IIB is S-selfdual, and thus the arguement could be
similar to the one used for going from M-theory to F-theory.

\vspace{3mm} Now looking at the IIA picture, M-theory on an
$11$-dimensional manifold $Y^{11}$ vs. F-theory \cite{Pope}
compactified on $Y^{11}\times S^1$. If we look at the Lagrangian
term \rref{estandard}, then it suggests that we can in fact
express the whole $Y^{11}\times S^1$-state by a state on $Y^{11}$,
valued not in the original target space, but in the loop space of
that space. Therefore, one could conjecture that the effective
action of F-theory compactified on $S^1$ can be computed in the
same index-theoretical way as the effective action of $M$-theory,
but instead of the $E_8$ and Rarita-Schwinger index terms, one
would substitute loop bundle versions of those indices. Now note
that the index of such operators should also be taken on loop
space, but it is well known that the relevant homotopical
information is contained at constant loops (this is well explained
in \cite{sell}), so integrating again over $Y^{11}$ seems
adequate.

\vspace{3mm} But now following \cite{sell}, replacing Dirac
operators by the corresponding Dirac operators on loop bundle
should correspond to replacing the $\hat{A}$-genus by the Witten
genus in the answer. Thus, this suggests a modification of the
method of \cite{DMW} to compare genuine $F$-theory partition
function obtained from its kinetic term, via its interpretation as
loop-bundle index on the spacetime $Y^{11}$ of M-theory, to the
elliptic `Witten genus' modification of the $K$-theoretical
partition function described in in detail above in this section.
We do not carry out this calculation here in detail, but propose
it as a concrete calculational esperiment which could be used to
test whether the field theory \cite{Pope} really has a consistent
compactification to M-theory.

\section{Remarks on signatures and supersymmetry}

\label{s6} Note however first that all the homotopy theory work
seriously described in this paper is done in Euclidean signature.
To discuss signatures seriously, we need to adapt our discussion
to manifolds with signatures. Here we simply point out the
relevance of signatures.

\vspace{3mm} As far as generalized cohomology with signatures, not
much has been done. Manifolds with signature typically cannot be
compact, so we must take cohomology with compact supports. But how
to take the signature into account in generalized cohomology? A
suggestive point is that $KO$-theory $KO^{p,q}\cong KO^{p-q}$
looks like it should be $KO$-theory of spacetime with signature
$(p,q)$. This, indeed, suggests a proposal: Ordinary cohomology,
$K$-theory and elliptic cohomology are all $\Z/2$-equivariant
generalized cohomology theories, which we can interpret as
generalized cohomology theories with a real form (\cite{hk}). Now
if $M$ is a manifold with signature, this makes the tangent bundle
$TM$ a $\Z/2$-equivariant bundle, where $\Z/2$ reverses signs of
purely time-like dimensions (this is not completely
Lorentz-invariant, but is so up to homotopy). Let us call this new
$\Z/2$-equivariant structure on the tangent bundle
$TM_{\epsilon}$. Then we can define, for a real-oriented
generalized cohomology $ER$, the signature-cohomology of $M$ as \(
ER^{k}_{c}(TM_{\epsilon}),\;\;k\in\Z \) where $c$ denotes compact
support. This is, of course, still a long way from working out all
the homotopy theory we have above at signatures, but it is a
start. We will develop the theory further elsewhere.

\vspace{3mm} We start with by looking at Clifford algebras in
twelve and eleven dimensions with various signatures. A discussion
on spinors in different dimensions and with various signatures can
be found in \cite{Kug}. In twelve dimensions, we are interested in
$(s,t)$ signatures, with $t=0,1,2,3$. The corresponding Clifford
algebras are isomorphic to the matrix algebras \bea (12,0) &:&
Mat_{32}({\mathbb H})
\\
(11,1) &:& Mat_{32}({\mathbb H})
\\
(10,2) &:& Mat_{64}(\R)
\\
(9,3)  &:& Mat_{64}(\R) \eea so that the pinor representations are
quaternionic in the first two cases and real in the last two
cases. For the spinor representation, one has to look at the even
Clifford algebra which is given by \( Cl(s,t)^{even} \cong
Cl(s-1,t)  ~~~{\rm for } s \geq 1. \) Then the even Clifford
algebras are given by \bea (12,0) &:& Mat_{16}({\mathbb H}) \oplus
Mat_{16}({\mathbb H})
\\
(11,1) &:& Mat_{32}(\C)
\\
(10,2) &:& Mat_{32}(\R)  \oplus Mat_{32}(\R)
\\
(9,3) &:& Mat_{32}(\C), \eea So one can have the following types
of spinors in twelve dimensions \bea (12,0) &:& {\rm Symplectic
~Majorana-Weyl}
\\
(11,1) &:& {\rm Majorana}
\\
(10,2) &:& {\rm Majorana-Weyl}
\\
(9,3) &:& {\rm Symplectic ~Majorana}. \eea

For the Lorentzian case, $(11,1)$, we have Majorana spinors. In
this case, one can try to form a supermultiplet for supergravity
formed out of 320 bosons and 320 fermions, but the gravitino and
the form sectors of the structure are incompatible \cite{Cartan}.
One can then ask whether one can construct supergravity theories
with other signatures in twelve dimensions. A general discussion
on this can be found in \cite{Sez}, and a proposal in the $(10,2)$
signature can be found in \cite{Hew} \cite{nishino}. Note that for
$(9,3)$ we can have symplectic-Majorana spinors, whose defining
relations for the charge-conjugation matrix $C$ and the gamma
matrices $\gamma^{\mu}$ are given by \( C^{T}=-C, ~~~
(\gamma^{\mu}C)^{T}=+\gamma^{\mu}C, ~~~
{\gamma^{\mu}}^{T}=-C^{-1}\gamma^{\mu} C. \) Some more discussion
on this from point of view of physics as well as mathematics will
be discussed seperately.

\vspace{3mm} Let us however make one final remark on a possible
significance of the signatures in connection with the IIA/IIB
duality. In the $(10,2)$ signature, the fiber is a Lorentzian
torus, which seems to break modularity. On the other hand, this
model seems forced if we want a physical version of the proposal
of \cite{Pope} (since signature $(9,3)$ does not contain $(10,1)$,
which is the physical signature of M-theory). This could be
consistent, since in type IIA, over which this sector of F-theory
is fibered, we indeed do not have manifest modularity.

\vspace{3mm} On the other hand, in type IIB theory, we need
manifest modularity, so it seems that physically, the
$(9,3)$-sector is required. However, now this sector of F-theory
cannot contain a physical $(10,1)$-M-theory, which again seems
consistent, as IIB theory does not seem to have a
$(10,1)$-M-theory dimensional expansion. It is possible that a
$(9,2)$ expansion is possible, and that this could in fact be the
correct physical signature for $\tilde{M}$-theory. This might be
not so unreasonable since there are versions of eleven-dimensional
M-theory in signatures $(1,10)$, $(2,9)$, $(5,6)$, $(6,5)$,
$(9,2)$, and $(10,1)$ \cite{Hull1}, \cite{Hull2}, \cite{Hull3}. In
fact in those theories, one already sees a difference between type
IIA and type IIB theories: while IIA allows for both $(10,0)$ and
$(9,1)$ signatures, type IIB allows for $(9,1)$ but not $(10,0)$.

\begin{appendix}
\section{Appendix: A brief review of topological modular forms.}
\end{appendix}

To make this paper more self-contained, we give here a very brief
review of the theories $tmf$ and $TMF$. This theory is due to Mike
Hopkins and Haynes Miller. All information necessary for our
purposes can be essentially found in \cite{HMA}. The main point is
this: in homotopy theory, it is convenient to consider
multiplicative (commutative associative) generalized cohomology
theories $E$ (also called spectra) which are $2$-periodic (in the
same way as $K$-theory), and are complex oriented, which means
that the generalized cohomology of the complex projective space is
of the form \beg{eap1}{E^*(\C P^{\infty})=E^*[[x]]} where $x$ is
the $E^*$-valued $1$-st Chern class of the universal line bundle
(equivalently, it suffices to say that such Chern class exists).
It then follows that all complex bundles have $E^*$-valued Chern
classes. In particular, one has
\beg{eap2}{E^*(\CP^{\infty}\times\C P^{\infty})=E^{*}[[1\otimes
x,x\otimes 1]]= E^*[[y,z]].} The multiplication $\C
P^{\infty}\times \C P^{\infty}\r \C P^{\infty}$ (classifying
tensor product of line bundles) then gives, via \rref{eap1},
\rref{eap2}, a map
$$E^{*}[[x]]\r E^{*}[[y,z]],$$
and the image of $x$ under this map is a series $F(y,z)$ called a
($1$-dimensional commutative) formal group law (abbr. FGL). Its
properties are
$$F(x,0)=x,$$
$$F(x,y)=F(y,x),$$
$$F(x,F(y,z))=F(F(x,y),z)).$$
Note that this looks like the properties of an analytic parametric
expansion of the multiplication in a $1$-dimensional commutative
Lie group. That is not very interesting, of course, since all such
groups are additive. Accordingly, even more generally, over a
field of characteristic $0$, all FGL's are isomorphic. However,
the essential point is that FGL's can be considered over any
commutative ring, and then this isomorphism statement is no longer
true. In fact, much information about a complex oriented
generalized cohomology theory can be deduced from its FGL. In
particular, the Lie group construction can be extended to
$1$-dimensional commutative algebraic groups, and this includes,
in addition to the additive and multiplicative group, also
elliptic curves. In fact, in the case of elliptic curves, it can
be extended even further, to generalized elliptic curves, which
only have multiplication defined in a {Zariski} neighborhood of
the identity. Details are irrelevant here (more precisely, are for
our purposes subsumed by what we shall say next). A
complex-oriented $2$-periodic spectrum whose FGL is isomorphic to
that of a generalized elliptic curve by a given isomorphism is
called an {\em elliptic spectrum}. (To be completely precise, it
is in fact convenient to add another condition that all
coefficient groups of elliptic spectra are in even dimensions.)

\vspace{3mm} Now algebraic geometers had long had to cope with the
fact that there is not, in the proper sense, a universal
generalized elliptic curve (for the same reason, there is also not
a universal elliptic cohomology theory), although the problem only
arises at the primes $2$, $3$. What there is, however, is the
Weierstrass curve, which is written, in affine coordinates, as
\beg{ewei}{y^2 +a_1xy+a_3y=x^3+a_2x^2+a_4x+a_6. } The coordinate
transformations allowed are
$$x=x^{\prime}+r,$$
$$y=y^{\prime}+sx^{\prime}+t.$$
Transformations for the $a_i$'s are easily deduced, but we do not
need to write them down for our purposes. The outcome is that we
obtain the pair
\beg{ea30}{(A,\Lambda)=(\Z[a_1,a_2,a_3,a_4,a_6],\Z[a_1,...a_6,r,s,t]).
} Here one should think of $r,s,t$ as free variables, i.e.
polynomial generators. When representing an actual
reparametrization of a generalized elliptic curve, they would take
values in the ring of definition of the curve. The main point is
that although every generalized elliptic curve is essentially a
Weierstrass curve (i.e. can be obtained by choosing the $a_1$'s
appropriately in an appropriate commutative ring), the pair
\rref{ea30} does not have the structure of coefficient rings of a
group scheme, thereby confirming that there indeed cannot be a
universal (generalized) elliptic curve. However, \rref{ea30}
satisfy the axioms of what is called an affine algebraic groupoid
(or, in homotopy theory, often Hopf algebroid). This proves that
there is a {\em Deligne-Mumford stack} of generalized elliptic
curves.

\vspace{3mm} Tensoring \rref{ea30} with $\Z[u,u^{-1}]$ where $u$
is an element of dimension $2$, we get
$(A[u,u^{-1}],\Lambda[u,u^{-1}])$. These graded rings can be
realized as coefficient rings of generalized elliptic spectra. Now
all of the difficulty of the construction of $tmf$ is contained in
the statement that the structure maps of \rref{ea30} (i.e. the
maps realizing its structure as an affine algebraic groupoid) can
be realized by maps of spectra (in particular generalized
cohomology theories). In fact, more is true, it can be generalized
by maps of $E_{\infty}$-ring spectra, which are commutative
associative ring cohomology theories in a particularly strong
sense. The spectrum $tmf$ is then defined as the homotopy inverse
limit of these structure maps, or equivalently of the system of
all $E_{\infty}$ elliptic spectra with respect to $E_{\infty}$
maps coming from morphisms of generalized elliptic curves.  This
construction was carried out in detail by Hopkins and Miller, and
recently much simplified by Jacob Lurie, using a remarkable
approach to algebraic geometry directly in the category of
$E_{\infty}$ ring spectra.

\vspace{3mm} Now just as there is no universal generalized
elliptic curve, there is no universal elliptic spectrum, so
accordingly, $tmf$ is not an elliptic spectrum. However, its
coefficient groups map to modular forms, and are called {\em
topological modular forms}. Not every form is a topological
modular form, and there are also topological modular forms which
are $0$ as ordinary modular forms. In particular, the discriminant
form $\Delta$ is not a topological modular form, but its $24$'th
power is. It is some times convenient to invert this $24$'th
power, thereby obtaining a $576$-periodic spectrum, which is
denoted by $TMF$.

\vspace{3mm} As we mentioned above, all the subtlety of $TMF$ is
at the primes $2,3$. When inverting $2,3$ in $TMF_{*}$, we obtain
simply ordinary modular forms:
$$\Z[1/6][g_2,g_3][\Delta^{-1}].$$
Completing at the prime $2$ (which is a slightly stronger
operation than localizing), the calculation of the homotopy groups
of $tmf$ is carried out in \cite{HMA}. There, one can in fact say
that there is a universal curve (with automorphisms). Its formal
group law is the Lubin-Tate law of height $2$. The curve can be
taken to be the curve $x^3+y^2+y=0$ over the $4$-element field
$\mathbb{F}_4$, and its group of rational points is
$\Z/3\times\Z/3$. We see there is a remarkable coincidence here
with modular forms of height $3$ over $\C$, which in fact plays a
major role in mathematics, but we do not need to consider this in
detail for the purposes of the present paper.




\begin{thebibliography}{99}
\bibitem{KS}
I.~Kriz and H.~Sati, {\it M Theory, type IIA superstrings, and
elliptic cohomology}, Adv. Theor. Math. Phys. {\bf 8} (2004) 345,
[{\tt arXiv:hep-th/0404013}].

\bibitem{KSB}
I.~Kriz and H.~Sati, {\em Type IIB string theory, S-duality, and
generalized cohomology}, to appear in Nucl. Phys. B, [{\tt
arXiv:hep-th/0410293}].

\bibitem{MW}
G.~Moore and E.~Witten, {\it Self duality, Ramond-Ramond fields,
and K-theory}, JHEP {\bf 05} (2000) 032, [{\tt
arXiv:hep-th/9912279}].

\bibitem{DMW}
E.~Diaconescu, G.~Moore and E.~Witten, {\it $E_8$ gauge theory,
and a derivation of K-Theory from M-Theory}, Adv. Theor. Math.
Phys. {\bf 6} (2003) 1031, [{\tt arXiv:hep-th/0005090}].

\bibitem{Wi1}
E.~Witten, {\it D-Branes and K-Theory}, JHEP {\bf 12} (1998) 019,
[{\tt arXiv:hep-th/9810188}].

\bibitem{Petr}
P. Ho\v{r}ava, {\it Type $IIA$ D-branes, K-theory, and matrix
theory}, Adv. Theor. Math. Phys. {\bf 2} (1999) 1373, [{\tt
arXiv:hep-th/9812135}].

\bibitem{HM}
V.~Mathai and H.~Sati, {\it Some relations between twisted
K-theory and $E\sb8$ gauge theory}, JHEP {\bf 03} (2004) 016,
[{\tt arXiv:hep-th/0312033}].

\bibitem{Vafa}
C. Vafa, {\it Evidence for F-theory}, Nucl. Phys. {\bf B469}
(1996) 403, [{\tt arXiv:hep-th/9602022}].

\bibitem{ell}
P. Hu and I. Kriz, {\em Conformal field theory and elliptic
cohomology}, to appear in Advances in Mathematics.

\bibitem{W-loop}
E. Witten, {\it The index of the Dirac operator in loop space}, in
\cite{LNM}, 161--181.

\bibitem{Sag}
S. Ferrara, R. Minasian and A. Sagnotti, {\it Low-energy analysis
of $M$ and $F$ theories on Calabi-Yau threefolds}, Nucl. Phys.
{\bf B474} (1996) 323, [{\tt arXiv:hep-th/9604097}].

\bibitem{Pope}
N. Khviengia, Z. Khviengia, H. Lu, and C. N. Pope, {\it Towads a
field theory of F-theory}, Class. Quant. Grav. {\bf 15} (1998)
759, [{\tt arXiv:hep-th/9703012}].

\bibitem{Allan}
A. Adams and J. Evslin, {\it The loop group of $E_8$ and K-theory
from $11d$}, JHEP {\bf 0302} (2003) 029, [{\tt
arXiv:hep-th/0203218}].

\bibitem{J1}
J.~Evslin, {\it From $E_8$ to F via T}, JHEP {\bf 0408} (2004)
021, [{\tt arXiv:hep-th/0311235}].

\bibitem{Hora}
P. Ho\v{r}ava and E. Witten, {\it Heterotic and type I string
dynamics from eleven dimensions}, Nucl. Phys. {\bf B460} (1996)
506, [{\tt arXiv:hep-th/9510209}].

\bibitem{flux}
E.~Witten, {\it On flux quantization in M-theory and the effective
action}, J. Geom. Phys. {\bf 22} (1997) 1, [{\tt
arXiv:hep-th/9609122}].

\bibitem{Seth}
M. B. Green and S. Sethi, {\it Supersymmetry constraints on Type
IIB supergravity}, Phys. Rev. {\bf D59} (1999) 046006, [{\tt
arXiv:hep-th/9808061}].

\bibitem{HT}
C. M. Hull and P. K. Townsend, {\it Unity of superstring
dualities}, Nucl. Phys. {\bf B438} (1995) 109, [{\tt
arXiv:hep-th/9410167}].

\bibitem{Var}
E. Witten, {\it String theory dynamics in various dimensions},
Nucl. Phys. {\bf B443} (1995) 85, [{\tt arXiv:hep-th/9503124}].

\bibitem{Gid}
S. B. Giddings, S. Kachru and J. Polchinski, {\it Hierarchies from
fluxes in string compactifications}, Phys. Rev. {\bf D66} (2002)
106006, [{\tt arXiv:hep-th/0105097}].


\bibitem{GG}
M. R. Gaberdiel and M. B. Green, {\it An $SL(2,\Bbb{Z})$ anomaly
in IIB supergravity and its F-theory interpretation}, JHEP {\bf
9811} (1998) 026, [{\tt arXiv:hep-th/9810153}].

\bibitem{Berg}
E. Bergshoeff, {\it Duality symmetries and the Type II string
effective action}, Nucl. Phys. Proc. Suppl. {\bf 46} (1996) 39,
[{\tt arXiv:hep-th/9509145}].

\bibitem{VW}
C. Vafa and E. Witten, {\it A one-loop test of string duality},
Nucl. Phys. {\bf B447} (1995) 261, [{\tt arXiv:hep-th/9505053}].

\bibitem{one}
M. J. Duff, J. T. Liu and R. Minasian, {\it Eleven dimensional
origin of string/string duality: A one loop test}, Nucl. Phys.
{\bf B452} (1995) 261, [{\tt arXiv:hep-th/9506126}].

\bibitem{GG2}
M. B. Green and M. Gutperle, {\it Effects of D-instantons}, Nucl.
Phys. {\bf B498} (1997) 195, [{\tt arXiv:hep-th/9701093}].

\bibitem{Roo}
M. de Roo, H. Suelmann and A. Wiedemann, {\it The supersymmetric
effective action of the heterotic string in ten dimensions}, Nucl.
Phys. {\bf B405} (1993) 326, [{\tt arXiv:hep-th/9210099}].

\bibitem{GSIII}
M. B. Green and J. H. Schwarz, {\it Supersymmetric dual string
theory (III). Loops and renormalization}, Nucl. Phys. {\bf B198}
(1982) 441.

\bibitem{Kir}
E. Kiritsis and B. Pioline, {\it On $R^4$ threshold corrections in
IIB string theory and $(p,q)$ string instantons}, Nucl. Phys. {\bf
B508} (1997) 509, [{\tt arXiv:hep-th/9707018}].

\bibitem{Das}
K. Dasgupta and S. Mukhi, {\it A note on low-dimensional string
compactifications}, Phys. Lett. {\bf B398} (1997) 285, [{\tt
arXiv:hep-th/9612188}].

\bibitem{pq}
J. H. Schwarz, {\it An $SL(2,Z)$ multiplet of type IIB
superstrings}, Phys. Lett. {\bf B360} (1995) 13; Erratum-ibid.
{\bf B364} (1995) 252, [{\tt arXiv:hep-th/9508143}].

\bibitem{Asp}
P. S. Aspinwall, {\it Some relationships between dualities in
string theory}, Nucl. Phys. Proc. Suppl. {\bf 46} (1996) 30, [{\tt
arXiv:hep-th/9508154}].

\bibitem{SchM}
J. H. Schwarz, {\it The power of M theory}, Phys. Lett. {\bf B367}
(1996) 97, [{\tt arXiv:hep-th/9510086}].

\bibitem{Gr}
M. B. Green, {\it Interconnections between type II superstrings, M
theory and $N=4$ supersymmetric Yang--Mills}, [{\tt
arXiv:hep-th/9903124}].

\bibitem{Piol}
 B. Pioline,
{\it A note on non-perturbative $R^4$ couplings}, Phys. Lett. {\bf
B431} (1998) 73, [{\tt arXiv:hep-th/9804023}].

\bibitem{Berk}
N. Berkovits, {\it Construction of $R^4$ terms in $N=2$ $D=8$
superspace}, Nucl. Phys. {\bf B514} (1998) 191, [{\it
arXiv:hep-th/9709116}].

\bibitem{MV}
D. R. Morrison and C. Vafa, {\it Compactifications of F-theory on
Calabi-Yau threefolds I}, Nucl. Phys. {\bf B473} (1996) 74, [{\tt
arXiv:hep-th/9602114}].

\bibitem{Mar2}
N. Marcus, {\it Composite anomalies in supergravity}, Phys. Lett.
{\bf B157} (1985) 383.

\bibitem{DFM}
 E.~Diaconescu, D.~Freed and G.~Moore,
{\it The M-theory 3-form and E8 gauge theory}, [{\tt
arXiv:hep-th/0312069}].

\bibitem{Sag2}
A. Sagnotti,{\it Open strings and their symmetry groups}, in
Cargese Summer Institute on Non-Perturbative Methods in Field
Theory, [{\tt arXiv:hep-th/0208020}].

\bibitem{AHS} M. Ando, M. J. Hopkins and N. P. Strickland
{\it Elliptic spectra, the Witten genus and the theorem of the
cube}, Invent. Math.  {\bf 146}  (2001) 595.

\bibitem{HMA}
M. J. Hopkins and M. Mahowald, {\it From elliptic curves to
homotopy theory}, preprint, [{\tt
http://hopf.math.purdue.edu/cgi-bin/generate?/Hopkins-Mahowald/eo2homotopy
}].

\bibitem{sell}
G. Segal, {\it Elliptic cohomology}, S\'{e}minaire Bourbaki, Vol.
1987/88, Ast\'{e}risque {\bf 161-162} (1988), Exp. No. {\bf 695}
(1989) 187.

\bibitem{LNM}
{\it Elliptic curves and modular forms in algebraic topology}, P.
S. Landweber (ed.), Lecture Notes in Math., 1326, Springer,
Berlin, 1988.

\bibitem{Stolz}
S.~Stolz and P.~Teichner, {\it What is an elliptic object?}, in
Topology, Geometry and Quantum Field Theory, U. Tillmann (ed.),
Cambridge University Press, 2004.

\bibitem{hk}
P. Hu and I. Kriz, {\it  Real-oriented homotopy theory and an
analogue of the Adams-Novikov spectral sequence}, Topology  {\bf
40} (2001) 317.

\bibitem{Kug}
T. Kugo and P. K. Townsend, {\it Supersymmetry and the division
algebras}, Nucl. Phys. {\bf B221} (1983) 357.

\bibitem{Cartan}
L. Castellani, P. Fre, F. Giani, K. Pilch and P. van
Nieuwenhuizen, {\it Beyond 11-dimensional supergravity and Cartan
integrable systems}, Phys. Rev. {\bf D26} (1982) 1481.

\bibitem{Sez}
I. Rudychev, E. Sezgin, and P. Sundell, {\it Supersymmetry in
dimensions beyond eleven}, Nucl. Phys. Proc. Suppl. {\bf 68}
(1998) 285, [{\tt arXiv:hep-th/9711127}].

\bibitem{Hew}
S. Hewson, {\it On supergravity in $(10,2)$}, [{\tt
arXiv:hep-th/9908209}].

\bibitem{nishino}
H. Nishino, {\it Supergravity in $10+2$ dimensions as consistent
background for superstring}, Phys. Lett. {\bf B428} (1998), [{\tt
arXiv:hep-th/9703214}].

\bibitem{Hull1}
 C. M. Hull,
{\it Timelike T-duality, de Sitter space, large $N$ gauge theories
and topological field theory}, JHEP {\bf 9807} (1998) 021, [{\tt
arXiv:hep-th/9806146}].


\bibitem{Hull2}
C. M. Hull, {\it Duality and the signature of space-time}, JHEP
{\bf 9811} (1998) 017, [{\tt arXiv:hep-th/9807127}].

\bibitem{Hull3}
C. M. Hull and R. R. Khuri, {\it Branes, times and dualities},
Nucl. Phys. {\bf B536} (1998) 219, [{\tt arXiv:hep-th/9808069}].

\end{thebibliography}
\end{document}